\documentclass{article}
\usepackage{spconf,amsmath,graphicx, amssymb}
\usepackage[nolist]{acronym}
\usepackage{mathtools, stmaryrd}
\usepackage{multirow}
\usepackage{rotating}
\usepackage{tikz}
\usepackage{url}
\usetikzlibrary{positioning}
\graphicspath{{./figs/}}

\title{DNN-based distributed multichannel mask estimation for speech enhancement in microphone arrays}
\twoauthors
  {Nicolas Furnon\thanks{This work was made with the support of the French National Research Agency, in the framework of the  project DiSCogs “Distant speech communication with heterogeneous unconstrained microphone arrays” (ANR-17-CE23-0026-01). Experiments presented in this paper were partially carried out using the Grid5000 testbed, supported by a scientific interest group hosted by Inria and including CNRS, RENATER and several Universities as well as other organizations (see https://www.grid5000), and using the EXPLOR centre, hosted by the University of Lorraine.},
  	Romain Serizel, Irina Illina}
	{Universit{\'{e}} de Lorraine, CNRS, Inria, Loria \\ F-54000 Nancy, France \\ \{firstname.lastame\}@loria.fr}
  {Slim Essid}
	{LTCI, T\'el\'ecom Paris, \\ Institut Polytechnique de Paris, \\Palaiseau, France \\
	slim.essid@telecom-paris.fr}
\begin{document}
\ninept
\begin{acronym}
\acro{mwf}[MWF]{multichannel Wiener filter}
\acro{sdw}[SDW-MWF]{speech distortion weighted multichannel Wiener filter}
\acro{mvdr}[MVDR]{minimum variance distortionless response}
\acro{gevd}[GEVD]{generalized eigenvalue decomposition}
\acro{nmf}[NMF-MWF]{non-negative matrix factorization}
\acro{tf}[TF]{time-frequency}
\acro{vad}[VAD]{voice activity detector}
\acroplural{vad}[VADs]{voice activity detectors}
\acro{danse}[DANSE]{distributed adaptive node-specific signal estimation}
\acro{mse}[MSE]{mean squared error}
\acro{wasn}[WASN]{wireless acoustic sensor network}
\acroplural{wasn}[WASNs]{wireless acoustic sensor networks}
\acro{doa}[DOA]{direction of arrival}
\acro{irm}[IRM]{ideal ratio mask}
\acro{ibm}[IBM]{ideal binary mask}
% NN
\acro{dnn}[DNN]{deep neural network}
\acroplural{dnn}[DNNs]{deep neural networks}
\acro{lstm}[LSTM]{long short-term memory}
\acro{gru}[GRU]{gated recurrent unit}
\acroplural{gru}[GRU]{gated recurrent units}
\acro{crnn}[CRNN]{convolutional recurrent neural network}
\acro{rnn}[RNN]{recurrent neural network}
\acroplural{rnn}[RNNs]{recurrent neural networks}
\acro{stft}[STFT]{short time Fourier transform}
\acro{rir}[RIR]{room impulse response}
\acroplural{rir}[RIRs]{room impulse responses}
\acro{snr}[SNR]{signal to noise ratio}
\acroplural{snr}[SNRs]{signal to noise ratios}
\acro{sar}[SAR]{source to artifacts ratio}
\acro{sir}[SIR]{source to interferences ratio}
\acro{sdr}[SDR]{source to distortion ratio}

%\acro{•}[•]{•}
%\acroplural{•}[•]{•}
\end{acronym}

\maketitle
\begin{abstract}
Multichannel processing is widely used for speech enhancement but several limitations appear when trying to deploy these solutions in the real world. Distributed sensor arrays that consider several devices with a few microphones is a viable solution which allows for exploiting the multiple devices equipped with microphones that we are using in our everyday life. In this context, we propose to extend the distributed adaptive node-specific signal estimation approach to a neural network framework. At each node, a local filtering is performed to send one signal to the other nodes where a mask is estimated by a neural network in order to compute a global multichannel Wiener filter. In an array of two nodes, we show that this additional signal can be leveraged to predict the masks and leads to better speech enhancement performance than when the mask estimation relies only on the local signals.
\end{abstract}
%
%%%%%%%%%%%%%%%%% INTRO %%%%%%%%%%%%%%%%%%%%%
\vspace{-4mm}
\section{Introduction}
\label{sec:intro}
Almost all voice-based applications such as mobile communications, hearing aids or human to machine interfaces require a clean version of speech for an optimal use. %Because of background noise surrounding the users, such a clean speech signal is not directly available and speech enhancement algorithms have been developed in order to reduce the noise.
Single-channel speech enhancement can substantially improve the speech intelligibility and speech recognition of a noisy mixture \cite{Gerkmann2012, Weninger2014}. However improvement with a single-channel filter is limited by the distortions introduced during the filtering operation. The distortion can be reduced in multichannel processing which exploits spatial information \cite{Frost1972, Vincent2018}.  %The multiplication of microphones in our daily life and the increase of the computational power of the devices they are embedded in raised attention to microphone arrays and multichannel signal processing. Given the \ac{doa}, data-independent beamformers \cite{Frost1972, Cox1987, VanVeen1988} can perform a spatial filtering on top of the spectral one. Data-dependent filters \cite{Buckley1987, Warsitz2007, Doclo2007}, which depend on the statistics of the source signals, reach higher performance than data-independent beamformers as they adapt to the acoustical properties of the auditory scene as well as to the statistical properties of the source signals \cite{Vincent2018}.
The \ac{mwf} \cite{Doclo2002} for example yields the optimal filter in the \ac{mse} sense and can be extended to a \ac{sdw} where the noise reduction is balanced by the speech distortion \cite{Doclo2007}.

Up to a certain point, the effectiveness of these algorithms increases with the number of microphones. More microphones can allow for a wider coverage of the acoustic scene and a more accurate estimation of the statistics of the source signals. In large rooms, or even in flats, this implies the need of huge microphone arrays, which, if they are constrained, can become prohibitively expensive and lacks flexibility. However, in our daily life,  with the omnipresence of computers, telephones and tablets, we are surrounded by an increased number of embedded microphones. They can be viewed as unconstrained ad hoc microphone arrays which are promising but also challenging \cite{Bertrand2015}. %Among them, the limited bandwidth and computational power available to send and process the signals between nodes.
%These filters are limited by the necessity that their microphones be synchronized or calibrated. Since a fusion center gathers all the measured signals, they do not scale well either as their computational complexity and bandwidth requirement increase exponentially with the number of microphones. In \acp{wasn}, no fusion centre is available and the several recording devices do not all share their signals and computational power. One way to cope with the reduced computational power is to develop algorithms where the cost is distributed over the nodes.
A \ac{danse} algorithm \cite{Bertrand2010}, where the nodes exchange a single linear combination of their local signals, was proposed for a fully connected microphone array. It was shown to converge to the centralized \ac{mwf} \cite{Bertrand2010a}. The constraint of a fully connected array can be lifted with randomized gossiping-based algorithms, where beamformer coefficients are computed in a distributed fashion \cite{Zeng2015}. Message passing \cite{Heusdens2012} or diffusion-based \cite{Oconnor2014} algorithms can increase the rather slow convergence rate of these solutions. Another way to exploit the broad covering of the acoustic field by ad hoc microphone arrays is to gather the microphones into clusters dominated by a single common source which can be estimated more efficiently \cite{Gergen2018}.

All these algorithms require the knowledge of either the \ac{doa} or the speech activity to compute the filters and are sensitive to signal mismatches \cite{Vorobyov2003} or detection errors \cite{Doclo2007}. Deep learning-based approaches have been proposed to estimate accurately these quantities through the prediction of a \ac{tf} mask \cite{Narayanan2013, Heymann2016, Perotin2018b} or of the spectrum of the desired signals \cite{Nugraha2016}. %Weninger et al.\ showed that a \ac{lstm} structured neural network performs better speech separation than previously used feed-forward \acp{dnn} by avoiding the vanishing gradient problem. Convolutional neural networks were introduced by Park and Lee in the context of speech enhancement \cite{Park2016} achieving comparable results than \acp{dnn} and \acp{rnn} but with much fewer parameters. A concatenation of a convolutional layer with a recurrent one is introduced in \cite{Zhao2018} where temporal as well as spectral structures are analysed by the network. %In \cite{Narayanan2013}, features are extracted from the mel-spectrogram of monaural signals and used to train \acp{dnn} in every frequency-channel which deliver the \ac{irm} of the mel-spectrogram. In \cite{Erdogan2015}, past as well as future temporal context is taken into account thanks to monodirectional and bidirectional \ac{lstm} cells. In \cite{Heymann2016}, the \ac{irm} is not directly applied on the mixture but used to compute the cross-power density matrices of the source signals in order to build a beamformer.
Although often used in a multichannel context, most of these solutions use single-channel data as input of their \acp{dnn}. Multichannel information was first taken into account through spatial features \cite{Jiang2014}, but can also be exploited using the magnitude and phase of several microphones as the input of a \ac{crnn} \cite{Adavanne2018, Chakrabarty2019}. This yields better results than single-channel prediction but combining all the sensor signals is not scalable and seems suboptimal because of the redundancy of the data. Coping with the redundancy, Perotin et al.\ \cite{Perotin2018a} combined a single estimate of the source signals with the input mixture and used the resulting tensor to train a \ac{lstm} \ac{rnn}.

In this paper, we consider a fully connected microphone array with synchronized sensors. This allows for using the \ac{mwf}-based \ac{danse} algorithm which was reported to achieve good speech enhancement performance \cite{Bertrand2010a}. Following the results shown by Perotin et al.\ \cite{Perotin2018a}, we take advantage of the \ac{danse} paradigm \cite{Bertrand2010a} by combining at each node one local signal with the estimations of the target signal sent by the other nodes. This uses the multichannel context for the mask estimation but avoids the redundancy brought by the signals of a same node. Additionally, this scheme takes advantage of the internal filter operated in \ac{danse} and reduces the costs in terms of bandwidth and computational power compared to a network combining all the sensor signals. %As this algorithm is distributed, it scales better than centralized solutions. It does not need any microphone calibration and the multichannel mask prediction allows for an accurate computation of the signals statistics.
%propose to replace the \ac{vad} required in DANSE \cite{Bertrand2010a} to estimate the speech and noise statistics by a \ac{tf} mask, which provides a finer estimation of the signals statistics. We estimate the \ac{tf} masks with \acp{dnn} at every node. In DANSE, each node sends a linear combination of its microphone signals, so-called compressed signal, to the other nodes in order to better estimate the local \ac{mwf}. We additionally provide this compressed signal to the \ac{dnn} by combining it with the signal observed at one microphone of the node. We compare the performance of two multichannel \acp{dnn}: one \ac{gru}  \cite{Cho2014} \ac{rnn} which is given a two dimensional concatenation of the observed and compressed signals; and one \ac{crnn} which is given a three dimensional concatenation of the same signals.

The paper is organised as follows. The problem formulation and \ac{danse} are described in Section \ref{sec:problem_formulation}. In Section \ref{sec:tf_estimation} we present our solution to estimate the \ac{tf} masks. The experimental setup is described in Section \ref{sec:setup} and results are discussed in Section \ref{sec:results} before we conclude the paper.
%This work is based on the \ac{danse} described in \cite{Bertrand11}. \ac{danse} is a two-staged \ac{mwf} where each node sends a linear combination of its microphone signals to the other nodes. Bertrand proves that this algorithm converges to the centralized \ac{mwf} solution, but reduces the required communication bandwidth. Like \ac{mwf}, it requires a \ac{vad} which we propose to substitute with a \ac{tf} mask. The frequency information contained in the \ac{tf} mask provides a finer estimation of speech and noise statistics. To estimate such masks, \cite{Heymann16} and \cite{Erdogan14} rely on single channel estimations or non causal models. We propose a \ac{crnn} to exploit information of several channels in an efficient manner.

%%%%%%%%%%%%%%%%% Problem formulation %%%%%%%%%%%%%%%%%%%%%
%\vspace{-2mm}
\section{Problem formulation}

\label{sec:problem_formulation}
\subsection{Signal model}
\label{subsec:signal_model}
We consider an additive noise model expressed in the \ac{stft} domain as $y(f, t) = s(f, t) + n(f, t)$ where $y(f, t)$ is the recorded mixture at frequency index $f$ and time frame index $t$. The speech target signal is denoted $s$ and the noise signal $n$. For the sake of conciseness, we will drop the time and frequency indexes $f$ and $t$. The signals are captured by $M$ microphones and stacked into a vector $\mathbf{y}~=~[y_{1}, ..., y_{M}]^T$. In the following, regular lowercase letters denote scalars; bold lowercase letters indicate vectors and bold uppercase letters indicate matrices.

\subsection{Multichannel Wiener filter}
\label{subsec:danse}
The \ac{mwf} operates in a fully connected microphone array. It aims at estimating the speech component $s_{i}$ of a reference signal at microphone $i$. Without loss of generality, we take the reference microphone as $i=1$ in the remainder of the paper. The \ac{mwf} $\mathbf{w}$ minimises the \ac{mse} cost function expressed as follows:
\begin{equation}
\label{eq:mse_cost}
J(\mathbf{w}) = \mathbb{E}\{|s_{1} - \mathbf{w}^H\mathbf{y}|^2\}.
\end{equation}
$\mathbb{E}\{\cdot\}$ is the expectation operator and $\cdot^H$ denotes the Hermitian transpose. The solution to (\ref{eq:mse_cost}) is given by
\begin{equation}
\label{eq:mwf_w}
\mathbf{\hat{w}} = \mathbf{R}_{yy}^{-1}\mathbf{R}_{ys}\mathbf{e}_1\,,
\end{equation}
with $\mathbf{R}_{yy} = \mathbb{E}\{\mathbf{y}\mathbf{y}^H\}$, $\mathbf{R}_{ys} = \mathbb{E}\{\mathbf{y}\mathbf{s}^H\}$ and $\mathbf{e}_1 = [1\; 0 \cdots 0]^T$. Under the assumption that speech and noise are uncorrelated and that the noise is locally stationary, $\mathbf{R}_{ys} = \mathbf{R}_{ss} = \mathbb{E}\{\mathbf{s}\mathbf{s}^H\} = \mathbf{R}_{yy} - \mathbf{R}_{nn}$ where $\mathbf{R}_{nn} = \mathbb{E}\{\mathbf{n}\mathbf{n}^H\}$. Computing these matrices requires the knowledge of noise-only periods and speech-plus-noise periods. This is typically obtained with a \ac{vad} \cite{Doclo2007, Bertrand2010a}.

The \ac{sdw} provides a trade-off between the noise reduction and the speech distortion \cite{Doclo2007}. The filter parameters minimise the cost function
\begin{equation}
\label{eq:cost_sdw}
J(\mathbf{w}) = \mathbb{E}\{|s_{1} - \mathbf{w}^H\mathbf{s}|^2\} + \mu \mathbb{E}\{|\mathbf{w}^H\mathbf{n}|^2\}\,,
\end{equation}
with $\mu$ the trade-off parameter. The solution to (\ref{eq:cost_sdw}) is given by
\begin{equation}
\label{eq:sdw_w}
\mathbf{\hat{w}} = \big(\mathbf{R}_{ss} + \mu\mathbf{R}_{nn}\big)^{-1}\mathbf{R}_{ss}\mathbf{e}_1.
\end{equation}
If the desired signal comes from a single source, the speech covariance matrix is theoretically of rank 1. Under this assumption, Serizel et al.\ \cite{Serizel2014} proposed a rank-1 approximation of $\mathbf{R}_{ss}$ based on a \ac{gevd}, delivering a filter that is more robust in low SNR scenarios and provides a stronger noise reduction. %We applied this decomposition to compute our filter coefficients.

\subsection{\ac{danse}}
In this section, we briefly describe the \ac{danse} algorithm under the assumption that a single target source is present. We consider $M$ microphones spread over $K$ nodes, each node $k$ containing $M_k$ microphones. The signals of one node $k$ are stacked in $\mathbf{y}_k~=~[y_{k,1}, ..., y_{k,M_k}]^T$.
As can be seen in (\ref{eq:mwf_w}), the array wide \ac{mwf} should be computed from all signals of the array, which can result in high bandwidth and computational costs. In \ac{danse}, only a single compressed signal $z_j$ is sent from node $j$ to the other nodes. So a node $k$ has $M_k + K -1$ signals, stacked in $\tilde{\mathbf{y}}_k = \left[ \mathbf{y}_k^T,~\mathbf{z}_{-k}^T\right]^T$,
%\[\tilde{\mathbf{y}}_k = \left[\begin{array}{c} \mathbf{y}_k  \\ \--\\
%\mathbf{z}_{-k}
%\end{array}
%\right]
%\]
where $\mathbf{z}_{-k}$ is a column vector gathering the compressed signals coming from the other nodes ${j\ne k}$. Replacing $\mathbf{y}$ by $\tilde{\mathbf{y}}_k$ and solving (\ref{eq:cost_sdw}) yields the \ac{danse} solution to the \ac{sdw}:
\begin{equation}
\label{eq:danse_w}
\mathbf{\tilde{w}}_k = \big(\mathbf{R}_{ss,k} + \mu\mathbf{R}_{nn,k}\big)^{-1}\mathbf{R}_{ss,k}\mathbf{e}_1\,,
\end{equation}
where $\mathbf{\tilde{w}}_k$, the filter at node $k$, can be decomposed into two filters as $\mathbf{\tilde{w}}_k~=~\left[ \mathbf{w}_{kk}^T,~\mathbf{g}_{k-k}^T\right]$. The first filter $\mathbf{w}_{kk}^T$ is applied on the local signals and $\mathbf{g}_{k-k}$ is applied on the compressed signals sent from the other nodes. The covariance matrices $\mathbf{R}_{ss,k}$ and $\mathbf{R}_{nn,k}$ are computed from the speech and noise components of $\mathbf{\tilde{y}}_k$. The compressed signal $z_{k}$ is computed as $z_{k} =  \mathbf{w}_{kk}^H\mathbf{y}_k$.
Bertrand and Moonen proved that this solution converges to the \ac{mwf} solution with $\mu = 1$, while dividing the bandwidth load by a factor $M_k$ at each node \cite{Bertrand2010a}. 

In this paper, we will focus on the batch-mode algorithm where the speech and noise statistics are computed based on the whole signal in order to focus on the interactions between the mask estimated by the \ac{dnn} and the \ac{mwf} filters.%This delivers a fully spatial processing since the filters are time-invariant, while delaying the study of the convergence hyperparameters to future work.

\section{Deep neural network based distributed multichannel Wiener filter}
\label{sec:tf_estimation}
\begin{figure}[t!]
	\centering
	\includegraphics[width=\linewidth]{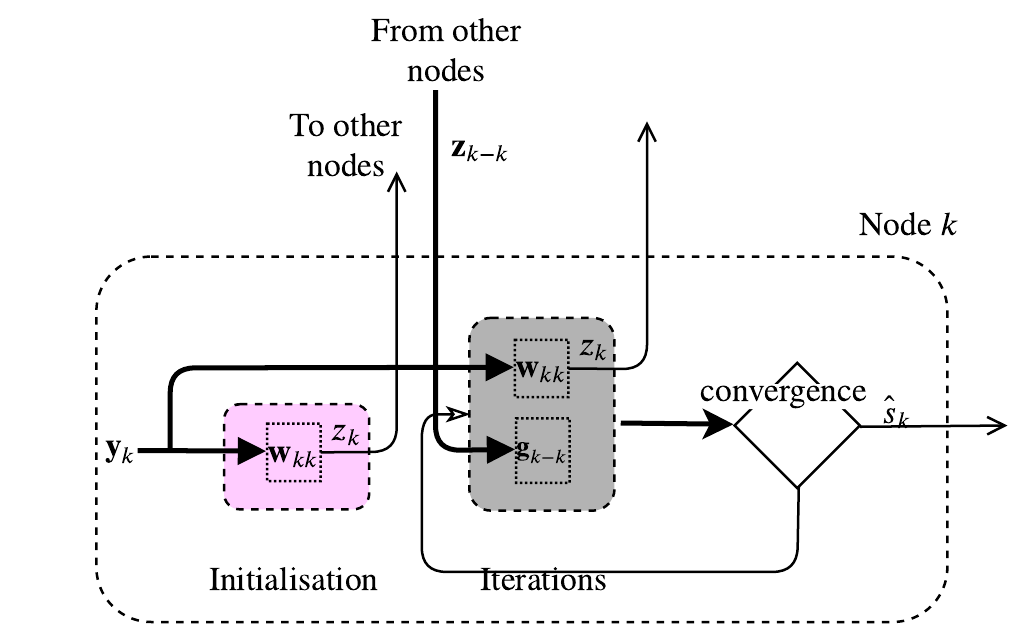}
	\caption{Block diagram of \ac{danse} principle. Bold arrows represent vectors, simple ones represent scalars.}
	\label{fig:danse}
\end{figure}

\begin{figure}[h]
	\centering
	\includegraphics[width=\linewidth]{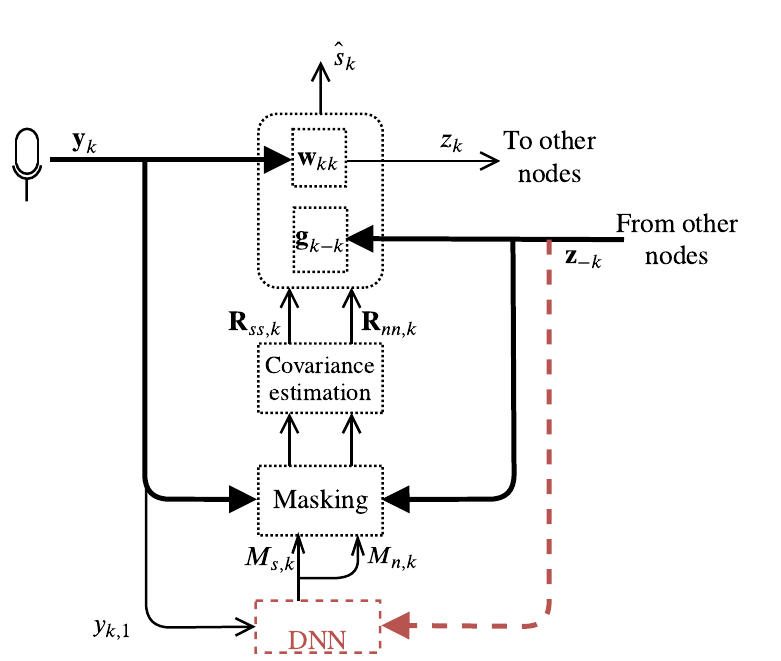}
	\caption{Expansion of the iterated step in Figure \ref{fig:danse}. Red parts are the modifications proposed to \ac{danse}. Bold arrows represent multichannel signals.}
	\label{fig:solution}
\end{figure}

Heymann et al.\ predicted \ac{tf} masks out of a single signal of the microphone array \cite{Heymann2016}. Perotin et al.\ \cite{Perotin2018a} or Chakrabarty and Habets \cite{Chakrabarty2019} included several other signals to improve the speech recognition or speech enhancement performance. We propose to extend these scenarios to the multi-node context of \ac{danse}.
In \ac{danse}, at node $k$, a single \ac{vad} is used to estimate the source and noise statistics required for both filters $\mathbf{w}_{kk}$ and $\mathbf{w}_{k}$. The first part of our contribution is to replace the \ac{vad} by a \ac{tf} mask predicted by a \ac{dnn}. Besides, since the compressed signals ${z}_{k}$ are sent from one node to the others, we also examine the option of exploiting this extra source of information by using it for the mask prediction. The schematic principle of \ac{danse} is depicted in Figure~\ref{fig:danse}. As it can be seen, an initialisation phase is required to compute the initial signal $z_k$.
We propose to do this with a first neural network. The second stage of \ac{danse} is represented in the greyed box in Figure \ref{fig:danse} and expended in Figure \ref{fig:solution}. Our second contribution is highlighted with the red arrow. It is to exploit the presence of $\mathbf{z}_{-k}$ at one node to better predict the masks with the \ac{dnn}. Several iterations are necessary for the filter $\mathbf{w}_{kk}$ to converge to the solution (\ref{eq:sdw_w}). In \ac{danse}, iterations are done at every time step. As we developed an offline batch-mode algorithm, we stopped the processing after the first iteration.
To analyse the effectiveness of combining $\mathbf{z}_{-k}$ with a reference signal to predict the mask, we compare our solution with a single-channel prediction, where the masks required for both initialisation and iteration stages are predicted by a single-channel model seeing only the local signal $y_{k,1}$.

%Finally, based on the results of Serizel et al. \cite{Serizel2014}, we used the rank 1 \ac{gevd} decomposition of the covariance matrices.

%The second is inspired by the \textit{U-Net} architecture first introduced in the biomedical imaging field \cite{Ronneberger2015} and later adapted to audio source separation \cite{Jansson2017}. It consists of an encoder made of ... \textit{to continue, depending on which architecture we introduce}.
We compare two different architectures for each of these schemes. The first architecture is a bidirectional \ac{lstm} introduced by Heymann et al.\ \cite{Heymann2016}.
When additional inputs are used with a \ac{rnn}, they are stacked over the frequency axis \cite{Perotin2018a}. Although this might deliver improved performance compared to the single-channel version, stacking it over the frequency axis is not efficient as many connections are used to represent relations between \ac{tf} bins that might not be related. That is why we propose a \ac{crnn} architecture which is more appropriate to process multichannel data. At each node, the compressed signals $\mathbf{z}_{-k}$ and the local reference signal $y_{k,1}$ are considered as separate convolutional channels.

During the training, in order to take into account the spectral shape of the speech, we weight the \ac{mse} loss between the predicted mask $\hat{\mathbf{m}}$ and the ground truth mask $\mathbf{m}$ by the \ac{stft} frame of the input $\mathbf{y}$, corresponding to the predicted frame. Both models are thus trained to minimise the cost function
\[
\mathcal{L}(\mathbf{m}, \hat{\mathbf{m}}) = E\{|(\mathbf{m} - \hat{\mathbf{m}})\cdot\mathbf{y}|^2\},
\]
where $E\{\cdot\}$ represents the empirical mean.

Lastly, since the filter $\mathbf{w}_{kk}$ is also applied on $\mathbf{z}_{-k}$, we use the \ac{gevd} of the covariance matrices to compute the \ac{mwf} of equation (\ref{eq:sdw_w}). Contrary to equation (\ref{eq:mwf_w}), this does not explicitly take the first microphone as a reference. It also assigns higher importance to the compressed signals, which is desirable since they are pre-filtered with potentially higher \acp{snr} than the local signals.

%%%%%%%%%%%%%%%%% SETUP %%%%%%%%%%%%%%%%%%%%%

\section{Experimental setup}
\label{sec:setup}

\subsection{Dataset}
\label{subsec:data}
Training as well as test data was generated by convolving clean speech and noise signals with simulated \acp{rir}, and then by mixing the convolved signals at a specific \ac{snr}. The anechoic speech material was taken from the clean subset of LibriSpeech \cite{Panayotov2015}. The \acp{rir} were obtained with the Matlab toolbox \textit{Roomsimove}\footnote{homepages.loria.fr/evincent/software/Roomsimove\_1.4.zip} simulating shoebox-like rooms.

In the training set, the length of the room was drawn uniformly as $l \in [3, 8]$\,m, the width as $w \in [3, 5]$\,m, the height as $h \in [2, 3]$\,m. Two nodes of four microphones each recorded the acoustic scene. The distance between the nodes was set to $1$\,m, the microphones being $10$\,cm away from the node centre. Each node was at least 1\,m away from the closest wall. One source of noise and one of speech were placed at $2.5$~m from the array centre. Both sources had an angular distance $\alpha \in [25, 90]^{\circ}$ relative to the array centre. The microphones as well as the sources were at the constant height of $1.5$~m. The \ac{snr} was drawn uniformly between $-5$\,dB and $+15$\,dB. The noise was white noise modulated in the spectral domain by the long term spectrum of speech. We generated $10,000$ files of 10 seconds each, corresponding to about 25 hours of training material.

The test configuration was the same as the training configuration but with restricted values for some parameters. The length of the room was randomly selected among $l \in \llbracket3, 8\rrbracket$\,m, the width among $w \in \llbracket3, 5\rrbracket$\,m, and the height was set to $h = 2.5$\,m. The angular distance $\alpha$ between the sources was randomly selected in $\alpha = \{25, 45, 90\}^{\circ}$. The noise was a random part of the third CHiME challenge dataset \cite{chime} in the cafeteria or pedestrian environment. We generated $1,000$ files representing about 2 hours of test material.

\subsection{Setup}
\label{subsec:setup}
All the data was sampled at 16~kHz. The \ac{stft} was computed with an FFT-length of 512 samples (32~ms), 50\% overlap and a Hanning window. 

Our \ac{crnn} model was composed of three convolutional layers with 32, 64 and 64 filters respectively. They all had $3\times3$ kernels, with stride $1\times1$ and ReLU activation functions. Each convolutional layer was followed by a batch normalization over the frequency axis and a maximum pooling layer of size $4\times1$ (along the frequency axis). The recurrent part of the network was a layer with 256 gated recurrent units, and the last layer was a fully connected layer with a sigmoid activation function.
The input data of both CRNN and RNN networks was made of sequences of 21 \ac{stft} frames and the mask corresponding to the middle frame was predicted. We trained them with the \mbox{RMSprop} optimizer \cite{rmsprop}.

%%%%%%%%%%%%%%%%% RESULTS %%%%%%%%%%%%%%%%%%%%%
\section{Results}
\label{sec:results}
We evaluate the speech enhancement performance based on the \ac{sar}, \ac{sir} and \ac{sdr} \cite{Vincent2006} computed with the \textit{mir\_eval}\footnote{https://github.com/craffel/mir\_eval/} toolbox. The performance reported corresponds to the mean over the $1,000$ test samples of the objective measures computed at the node with the best input \ac{snr}. We also report the 95\% confidence interval.

The \ac{gevd} filter does not explicitly take one sensor signal as the reference signal to minimise the cost function, but a projection of the input signals into the space spanned by the common eigenvectors of the covariance matrices. Because of that, the objective measures computed with respect to the convolved signals did not give results that were coherent with perceptual listening tests performed internally on random samples. Indeed, differences between the enhanced signal and the reference signal are interpreted as artefacts whereas they are due to the decomposition of the input signals into the eigenvalue space of the covariance matrices. Therefore, we compute the objective measures using the dry (source) signals as reference signals. This decreases the \ac{sar} because the reverberation is then considered as an artefact but the comparison between methods correlates more with the perceptual listening tests.

We present the objective metrics for the different approaches in Table~\ref{tab:res_oim}. In this table, single node filters are referred to as MWF (upper part of the table) and distributed filters as DANSE (lower part of the table). For each filter, the architecture used to obtain the masks is indicated between parenthesis. RNN refers to Heymann's architecture and CRNN to the network introduced in Section \ref{subsec:setup}. The subscript of the network architecture indicates the channels considered at the input. The results obtained with the single-channel \ac{dnn} models are denoted with "SC". When the compressed signals $\mathbf{z}_{-k}$ were used as additional input to the \ac{dnn} to predict the mask of the second filtering stage, models are denoted with "MC". Additionally, we report the number of trainable parameters of each model in Table \ref{tab:params}.

\subsection{Oracle performance}
\label{subsec:oracle_se}
The \ac{vad} gives information about the speech-plus-noise and noise-only periods in a wide-band manner only, whereas a mask gives spectral information that enables a finer estimation of the speech and noise covariance matrices. This additional information is translated into an improvement of the speech enhancement performance with both types of filters (MWF and DANSE). In the following section, we analyse whether this conclusion still holds when the masks are predicted by a neural network.

\subsection{Performance with predicted masks}
\label{subsec:mc_mp}

\begin{table}[h]
	\setlength\tabcolsep{6pt}
	\centering
	\renewcommand{\arraystretch}{1.2}
	\begin{tabular}{|c|ccc|}
		\hline
		(dB)  & SAR & SIR & SDR \\
		\hline
		\hline
		MWF (oracle \ac{vad}) & 2.4$\pm$0.3		&	24.7$\pm$0.3	& 2.3$\pm$0.3	\\
		MWF (oracle mask) & 	4.0$\pm$0.3	&	26.7$\pm$0.3	& 3.9$\pm$0.3	\\
		\hline
		MWF (RNN) & 	3.4$\pm$0.3	&	25.1$\pm$0.4	& 3.3$\pm$0.3	\\
		MWF (CRNN) & 	3.3$\pm$0.3	&	25.1$\pm$0.4	& 3.2$\pm$0.3	\\
		\hline
		\hline
		DANSE (oracle VAD) &2.6$\pm$ 0.3 & 25.2$\pm$ 0.3 & 2.6$\pm$ 0.3\\
		DANSE (oracle mask) & 4.8$\pm$ 0.3 & 27.6$\pm$ 0.3 & 4.8$\pm$ 0.3 \\
		\hline
		DANSE (RNN$_{\mathrm{SC}}$)  & $4.0 \pm 0.3$ 	& 26.0$\pm$0.4   & $4.0\pm0.3$   \\
		DANSE (CRNN$_{\mathrm{SC}}$)   & $4.0\pm0.3$ 	& $26.0\pm0.4$   & $4.0\pm0.3$   \\
		\hline
		DANSE (RNN$_{\mathrm{MC}}$)  & $4.1\pm 0.3$ 	& $26.1\pm0.4$   & $4.0\pm0.3$   \\
		DANSE (CRNN$_{\mathrm{MC}}$)  & 4.7$\pm$0.3 	& 27.4$\pm$0.3   & 4.6$\pm$0.3   \\
		\hline

	\end{tabular}
\caption{Speech enhancement results in dB with oracle activity detectors and predicted ones.}
\label{tab:res_oim}
\end{table}

First, replacing the oracle \ac{vad} by masks brings significant improvement in terms of all objective measures. This confirms the idea that \ac{tf} masks are better activity detectors than \acp{vad}, even oracle ones. Second, the objective measures corresponding to the output signals of DANSE filters are always better than those of the MWF filters. This confirms the benefit of using the \ac{danse} algorithm. Although these differences are not high, increasing the number of nodes and the distance between them might enhance the utility of the distributed method.

From the results in Table \ref{tab:res_oim}, there is no clear advantage of using a \ac{crnn} over using a \ac{rnn} in the single channel case. Indeed, the objective measures of RNN$_{\mathrm{SC}}$ and CRNN$_{\mathrm{SC}}$ match in all points. In the multichannel case, the performance of the \ac{rnn}-based approach does not increase. This tends to confirm that the \ac{rnn} is not able to efficiently exploit multichannel information. Since the \ac{rnn} delivered good results in the single-channel scenario, this leads to the conclusion that stacking multichannel input on the frequency axis is not appropriate. In addition, as shown in Table \ref{tab:params}, the number of parameters of the \ac{rnn} almost doubles when a second signal is used, whereas it barely increases for the \ac{crnn}. This is due to the convolutional layers of the \ac{crnn} which can process multichannel data much more efficiently than recurrent layers.

The \ac{crnn} solution can exploit the multichannel inputs efficiently and the performance increases for all metrics. The biggest improvement is obtained for the \ac{sir}. Indeed, one of the main difficulties for the models is to predict noise-only regions, because of people talking in the noise CHiME database. Since the compressed signals are pre-filtered, they contain less noise and they are less ambiguous. This makes it easier for the model to recognize noise-only regions, without degrading its predictions of speech-plus-noise regions.

\begin{table}[t!]
	\centering
	\renewcommand{\arraystretch}{1.1}
	\begin{tabular}{|c|c|}
		\hline
		Model & Number of parameters \\
		\hline
		RNN$_{\mathrm{SC}}$ & $1,717,773 $            \\
		CRNN$_{\mathrm{SC}}$  & $911,109$  \\
		\hline
		RNN$_{\mathrm{MC}}$ &$2,244,109$\\
		CRNN$_{\mathrm{MC}}$  &$911,397$  \\
		\hline
	\end{tabular}
	\caption{Number of trainable parameters of the neural networks.}
	\label{tab:params}
\end{table}
%\vspace{-4mm}
\section{Conclusion and future work}
\label{sec:conclusion}
We introduced an efficient way of estimating masks in a multi-node context. We developed multichannel models combining an estimation of the target signals sent by the other nodes with a local sensor. This proved to better predict \ac{tf} masks, which led to higher speech enhancement performance that outperformed the results obtained with an oracle \ac{vad}. A \ac{crnn} was compared to a \ac{rnn} and the \ac{crnn} could exploit much better the multichannel information. In addition, the \ac{rnn} architecture is limited by its number of parameters, especially if the number of nodes had to increase. In such scenarios, the difference between single-channel and multichannel models performance might be even more important but this still has to be explored. To attain performance closer to the oracle ones, several options are possible. First, the rather simple architectures that were used could be replaced by state-of-the art architectures. Besides, given the increase in performance when the target estimation is given, it would also be interesting to additionally give the noise estimation at the input of the models. %Lastly, an adaptive version seems necessary to dedicate this algorithm to real-time applications.

%\begin{figure}
%\begin{tikzpicture}[x=0.1cm,y=0.1cm]
%	\node[inner sep=0pt] (input) at (0,0)
%	{\includegraphics[width=.25\textwidth]{figs/input_stft.png}};
%		\node[align=left] at (7, -5.5) {\footnotesize 21 x 256};
%	\node[draw,align=center, text width=0.25\textwidth](conv) at (0,-25) { \small 2D Conv 3x3, 32 filters \\ ReLu, BatchNormalization \\ MaxPooling 1 x 4 \\ 2D Conv 3x3, 64 filters\\ ReLu, BatchNormalization \\ MaxPooling 1 x 4 \\ 2D Conv 3x3, 64 filters \\ ReLu, BatchNormalization \\ MaxPooling 1 x 4};
%			\node[align=left] at (7, -52.5) {\footnotesize 15 x 4 x 64};
%	\node[draw](reshape) at (0,-60) {Reshape};
%			\node[align=left] at (7, -66.5) {\footnotesize 15 x 256};
%	\node[draw](rnn) at (0,-75) {\small GRU, 256 units};
%			\node[align=left] at (7, -81.5) {\footnotesize 1 x 256};
%	\node[draw](ff) at (0,-90) {\small Fully connected layer, 257 units};
%	\node[inner sep=0pt] (output) at (0,-100)
%	{\includegraphics[width=.25\textwidth]{figs/output_mask.png}};
%	\draw[->] (input) edge (conv) (conv) edge (reshape) (reshape) edge (rnn) (rnn) edge (ff) (ff) edge (output);
%\end{tikzpicture}
%\label{fig:crnn}
%\caption{Structure of the \ac{crnn}for the mask estimation}
%\end{figure}

\vfill\pagebreak
\bibliographystyle{IEEEbib}
\bibliography{strings,refs}
\label{sec:refs}

\end{document}